# Event-based Signal Processing for Radioisotope Identification

Xiaoyu Huang[1], Edward Jones[2], Siru Zhang[3], Steve Furber[4], *Fellow, IEEE*, Yannis Goulermas[5],
Edward Marsden[6], Ian Baistow[7], Srinjoy Mitra[8], Alister Hamilton[9]
[1, 8, 9]University of Edinburgh, UK, [2, 4]University of Manchester, UK, [3, 5]University of Liverpool, UK, [6, 7]Kromek Group plc, UK
E-mail: {[1]xiaoyu.huang, [8]srinjoy.mitra, [9]alister.hamilton}@ed.ac.uk, {[2]edward.jones-3, [4]steve.furber}@manchester.ac.uk,
{[3]siru.zhang, [5]j.y.goulermas}@liverpool.ac.uk, {[6]ed.marsden, [7]ian.baistow}@kromek.com

*Abstract*—this paper identifies the problem of unnecessary high power overhead of the conventional frame-based radioisotope identification process and proposes an event-based signal processing process to address the problem established. It also presents the design flow of the neuromorphic processor.

## I. Introduction

The detection and identification of radioisotope material play an important role in national security; to help counter the terrorist threat of dirty bombs (radiological dispersal devices) as well as potential nuclear devices. Radiation detectors are an important technology to determine the type and dose of radiation. The motivation of this research is to develop building blocks which will help drive down the power consumed by intelligent radiation detectors in order to help reduce their size and improve the capabilities for application in mobile scenarios.

The power consumption of sensor-supporting electronics for signal amplification and processing currently limits the capabilities of mobile detectors for nuclear materials. Our project partner company Kromek has developed a range of handheld gamma and neutron detector, such as the D3S, for Defense Advanced Research Projects Agency (DARPA) as part of the Securing the Cities program. With a 1450 *mAh* battery and 100 *mA* processor the device has a typical running time of around 12 hours, long enough for such a device to be carried by a police officer during a typical shift but limiting the applications beyond this.

## II. Background

### A. The Problem

As illustrated in Fig. 1a, the conventional frame-based radioisotope identification (ID) process involves the following steps:

*1) Scintillation:* the invisible gamma photon interacts with the scintillation material causing the emission of one or several photons in the visible range.

*2) Photon detection:* light photons are detected by the photodetector and continuous voltage signals are produced, see Fig. 1a.

*3) Analogue to digital conversion (ADC) and integration:* for each detection event the analogue electrical signal, which may represent the sum of several scintillation events from a single incident gamma photon, is converted to a digital signal, and integrated over time to give a value for each event proportional to the energy of the incident gamma photon.

*4) Histogram generation:* the counts of gamma events at each energy channel are accumulated to generate an energy histogram or spectrum for a given time window.

*5) Radioisotope Identification:* each target radioisotope has a signature histogram making radioisotope identification a histogram classification problem. Many algorithms can be applied to this kind of task, notably the Poisson-Clutter Split (PCS) algorithm [1] has been applied to the radioisotope identification histogram classification task in the past.

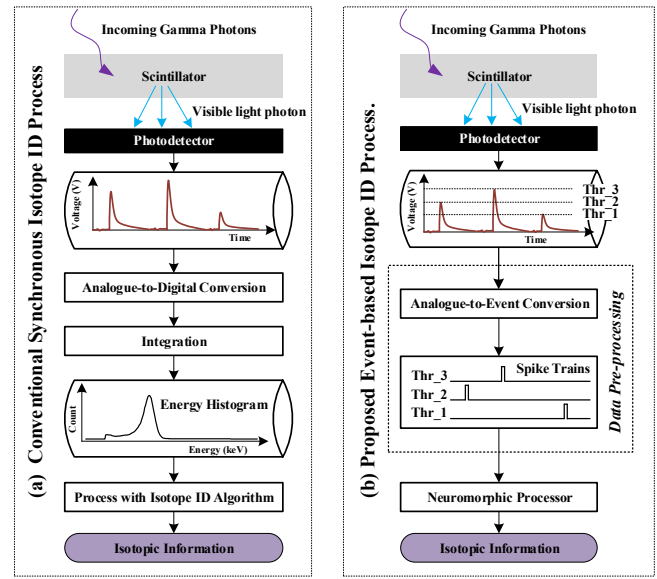

Fig. 1. Conventional Frame-based Process and Proposed Event-based Process

In this frame-based algorithm described in Fig. 1, the ADC and integration computational units are normally implemented using the synchronous techniques and the energy histogram needs to be generated for each frame. The main problem with this approach is that these data processing units continuously consume power, even when the event arrival rates are very low or zero.

### B. Proposed Method

We propose an event-driven algorithm that could process signals asynchronously. This means that processing is initiated only when detection events take place, giving the potential for improved energy efficiency. Fig. 1b shows the proposed isotope ID process where the continuous analogue voltage signal is converted into 'events' which are then processed by neuromorphic hardware for isotope identification. The details of neuromorphic processor design are as presented in section III.

## III. METHODOLOGY

Neuromorphic hardware is an emerging field of research that seeks to design computing devices with brain-inspired architectures. Emulating biological neural networks in silicon hardware, neuromorphic architectures are expected to solve complex problems with low power consumption in an event-driven manner [2]. The most established application domain for event-based models is that of computer vision where event-based cameras and processing can be used such that only per-pixel brightness changes are measured asynchronously, avoiding the redundancy seen in traditional frame-based approaches [3][4]. Due to their spatiotemporal nature and intermediate level of abstraction between biological plausibility and the Artificial Neural Networks (ANNs) of machine learning, Spiking Neural Networks (SNNs) are a popular model for the implementation of these low power event-based systems [5]. Fig. 2 illustrates the proposed design flow of an event-based neuromorphic processor for radioisotope identification.

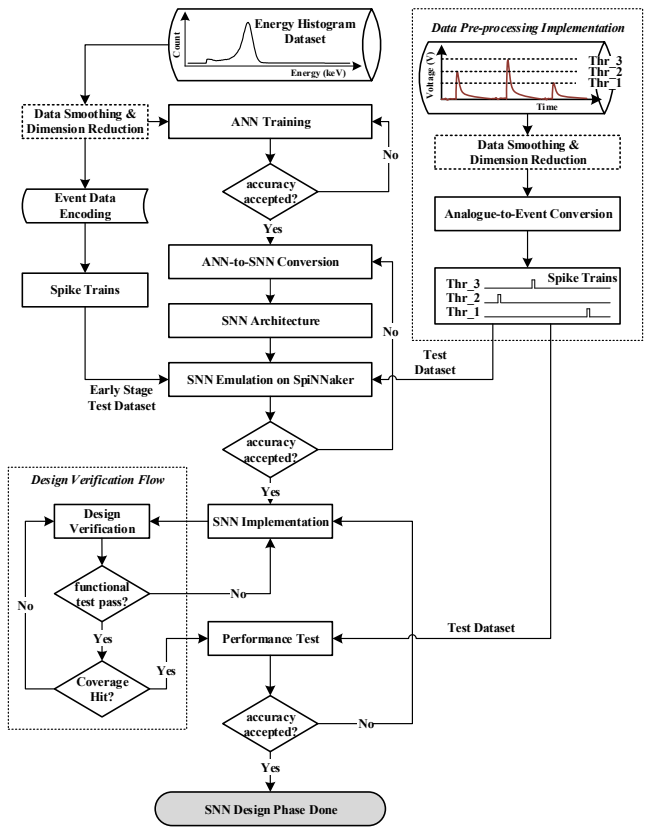

Fig. 2. Design Flow of the Event-based Neuromorphic Processor

### A. Dataset

As shown in Fig.3, the dataset was recorded using two different measurement setups, with and without a Polymethyl Methacrylate (PMMA) phantom, which is designed to represent the upper torso of a user. Spectra for each radioisotope source under test were recorded every second for 120 seconds. Subsequent measurements were made at varying distances, which are 10 *cm*, 25 *cm*, 50 *cm*, 1 *m* and 1.5 *m*, with each source.

*1) Analogue Electrical Signal:* Fig. 4 shows a screen shot for the Silicon Photomultiplier (SiPM) anode output signal, collected across a 50 *Ω* load resistor. The gamma event shown has several thousand photons. The SiPM is an avalanche device with a gain of about 106.

*2) Energy Histogram Dataset:* There are 4096 channel bins in the raw dataset, which is based on the 12-bit ADC. They are calibrated into 3238 energy bins. Fig. 5 illustrates an energy histogram of the radioisotope source Americium-241 during 120 seconds at the distance of 10 *cm* with PMMA phantom.

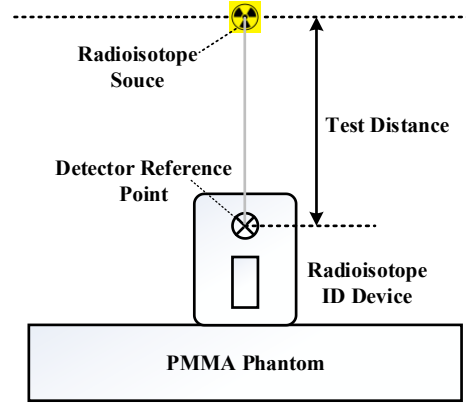

Fig. 3. Test Geometry for Photon Alarm

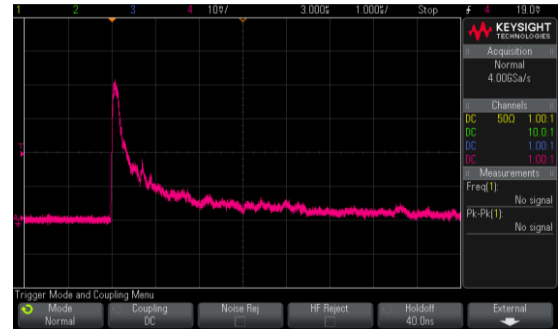

Fig. 4. Analogue Electrical Signal

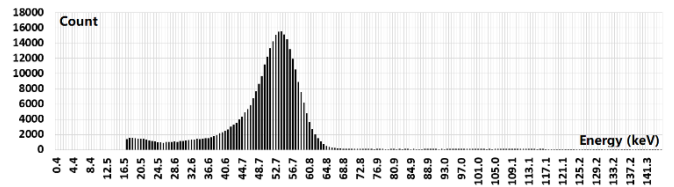

Fig. 5. Energy Histogram of radioisotope source Americium-241

### B. Data Pre-processing

*1) Data Smoothing and Dimension Reduction:* the performance of the classifier could be afftected by the variations from Poisson count statistics [6]. Data smoothing techiniques, such as local regression smoother and wavelet smoother, can be applied to remove the noise to maximise the performance under the limited hardware resources. To reduce the complexity of the design and save power, the dimension reduction techniques such as Poisson PCA can be utilised to remove the redundant features of the dataset [7] [8].

*2) Analogue-to-event Conversion:* the multiple threshold technique in [9] can be used to convert the analogue representation of the single Gamma photon energy level into the event domain. Each energy threshold can be considered as a signature. In this event domain, the signature of the target radioisotope is manifested in form of spike trains. The SNN is trained to recognize those spike trains to provide the radioisotopic classification.

*3) Event Data Encoding:* in parallel with the implementation of the data pre-processing, the spike trains of the radioisotope manifestation can also be encoded based on energy histograms from the frame-based process. This process extracts the information at time, encodes the data into spikes and stretches them into event data streams. It can be utilized as test dataset for the SNN emulation at very early stage, which can speed up the development process prior to the analogue-to-event implementation being available.

*C. Artificial Neural Network (ANN) Training*

As mentioned in the Section II – A4, the energy histogram generated by accumulating counts for each channel over a certain integration time in the frame-based process can be used for radioisotope identification. ANNs can be trained to recognize the energy histogram of given radioisotopes and provide the isotope classification. The reasons for choosing ANNs for this task are 1) ANNs perform well at classification tasks in analogous computer vision applications; 2) ANN training methods and development tools are well established and the development process can proceed much faster than designing an SNN from scratch; 3) Tools and techniques for ANN-to-SNN conversion are becoming available [10].

*D. ANN-to-SNN Conversion*

SNNs represent neural activity as a series of spikes over time and so are inherently temporal, ideal for an event-based task such as radioisotope identification. Eligibility propagation is a promising method for training SNNs with performance comparable to that of back-propagation training in ANNs [5] however we chose to train ANNs via back-propagation on histogram data and to convert the trained ANNs to SNNs because of how much more establish this path is. For ANN-to-SNN conversion we use a modified version of the SNN toolbox as developed by Rueckauer et al. [10]. The neuron model used is a standard PyNN current-based leaky integrate-and-fire (*LIF_curr_exp*) neuron model [11] with the default parameters given in the SNN toolbox. The activation values seen in the ANN models are represented as average spike rates in the SNN. The input histograms are converted to the rates of Poisson sources corresponding the energy channel of the histogram.

*E. Neural Network Architecture*

A simple four layer convolutional neural network was designed for the task of radioisotope identification using the Keras library [12].

A diagram of the neural network architecture used is shown in Figure 6. It is made up of an input layer the size of the calibrated input (3238), two one-dimensional convolutional layers and a dense output layer. The ANN is trained on the continuous histogram data and this input is transformed into the spiking domain by using the continuous values as the rates of Poisson spike sources in the spiking model on SpiNNaker.

*F. SNN Emulation on SpiNNaker*

SpiNNaker is a neuromorphic hardware platform designed for the simulation of large-scale spiking neural networks at speeds close to biological real-time [13]. In this project SpiNNaker is used to run SNNs that are the product of ANN-to-SNN conversion using the method described by Rueckauer et al. in [10]. The energy efficiency and scale of SpiNNaker systems (notably the 1 million core machine in Manchester, UK) allows for rapid prototyping of network architectures and for parallel optimization algorithms such as a genetic algorithms to be used to tune network parameters.

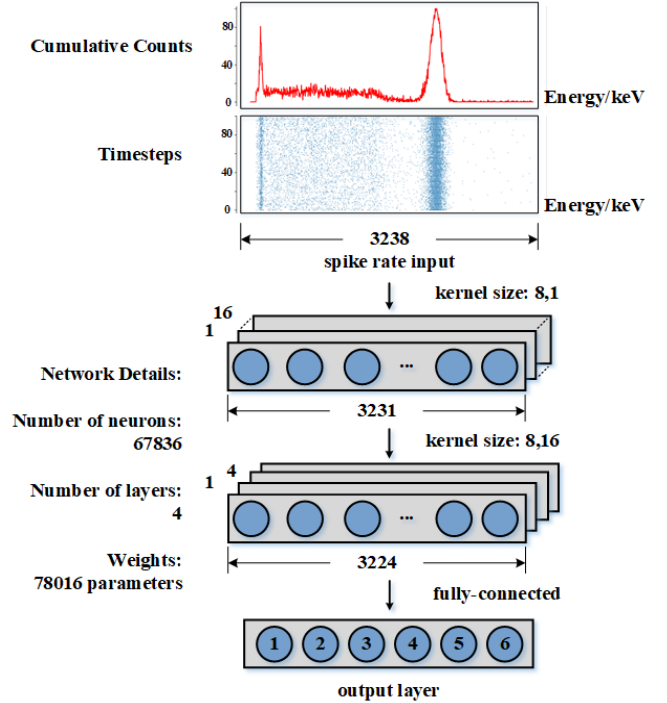

Fig. 6. An illustration of the SNN architecture. Input energy histograms are encoded to spike trains and then fed into SNN. The input is 3238, and the number of neurons in the remaining layers is given by 51696-12896-6. All neurons in the network are spiking neurons.

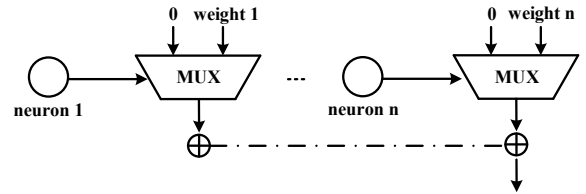

Fig. 7. Synaptic microarchitecture [14]

*G. SNN Implementation in Hardware*

Once verified and optimized through model simulation on SpiNNaker, the SNN architecture is then implemented on a Field-Programmable Gate Array (FPGA).

In many SNN implementation methods, high precision (e.g. 32-bit) neurons and synapses are applied to provide continuous derivatives and support incremental changes to network state [15][16]. However [14] reported that such high precision is not necessary due to the redundancy of networks. Its experimental results on the MNIST dataset also demonstrated that both the refractory period and alpha-function-shaped post-synaptic current are not compulsory to achieve high performance inference. Therefore in this work, this fashion is followed. Besides as shown in Fig. 7, a digital manner of the post-synaptic current input is applied to save power and cost. The $x(n)$ represents either the presence (1) or absence (0) of a spike.

### H. Design Verification and Performance Test

Due to the long development path, a tiny mismatch in the functionality in the initial design implementation may cause a large potential loss in performance in the final implementation. The debug process in this workflow could be very expensive and could significantly increase the time-to-market. For these reasons, design verification during implementation is crucial.

The implementation should be verified by functional and random tests at the block level, such as in the LIF Neuron Model block, and also through integration tests at the system level. The verification process is complete when all the tests are passed and both the functional and code coverage are closed.

Once the functionality of the implementation is fully verified, the performance tests with test dataset as stimuli are then run to check whether the accuracy of the radioisotope identification is as anticipated.

## IV. RESULTS

An ANN model was built based on the architecture outlined in Figure 6. The weights were trained by backpropagation on a cut-down dataset of 6 classes: 5 industrial radioisotopes ($^{241}$Am, $^{133}$Ba, $^{60}$Co, $^{137}$Cs, $^{152}$Eu) and background. The resulting ANN gave a testing accuracy of 100% on a test set of 100 examples and 99.94% on 1692 test examples. The dataset used for ANN training was continuous histogram data that was transformed into Poisson rates for SNN evaluation in the spiking domain. The SNN returned an accuracy of 85% on a test set of 100 examples.

## V. DISCUSSION

The experiments carried out here are limited in both their dataset and the conversion accuracy. A degradation in performance is seen in the conversion of the ANN to SNN most likely due to sub-optimal SNN neuron model parameters and the limited presentation duration: how long the inputs are applied to the network.

The experiments carried out were on a limited dataset in which the test data were statistically very similar to the training data. In future work we will use more data and will look at synthesising data to make approaches robust to noise, variations in background radiation, source intensity and distance between source and detector.

## VI. CONCLUSION

The power overhead associated with a frame-based radioisotope identification process has been identified and a low power event-based substitute has been proposed. This paper has detailed the design flow, discussed the relevant implementation details of this approach and presented some preliminary results.

At an early stage, the results represent a proof of principle in the application of SNNs converted from ANNs to the task of radioisotope identification.

To achieve the ultimate goal of an ultra-low power design, considerations in data pre-processing, SNN architecture and the hardware implementation method need to be taken. Therefore our further work will focus on more efficient data pre-processing techniques, the use of optimization methods to tune the SNNs to be more robust and applicable to real-world application and their power efficient hardware implementation methods.

ACKNOWLEDGMENT

This project is funded by US Defense Threat Reduction Agency (DTRA) and Kromek Group plc, UK.